\documentclass[preprint2]{aastex}

\shorttitle{Fiber measurements with enhanced reliability}
\shortauthors{Schmoll \& Roth \& Laux}

\begin{document}

\title{Statistical Test of Optical Fibers for use in PMAS, the Potsdam Multi-Aperture Spectrophotometer}

\author{J. Schmoll\altaffilmark{1}}
\affil{Astronomical Instrument Group, Dept. of Physics, Rochester Buildg., University of Durham, South Road, Durham DH1 3LE, United Kingdom, Email: jurgen.schmoll@durham.ac.uk}
\author{M.M. Roth}
\affil{Astrophysikalisches Institut Potsdam, An der Sternwarte 16, D-14482 Potsdam, Germany \\ Email: mmroth@aip.de}
\author{U. Laux}
\affil{Th\"uringer Landessternwarte Tautenburg, Sternwarte 5, D-07778 Tautenburg, Germany \\ Email: laux@tls-tautenburg.de}

\altaffiltext{1}{Formerly Astrophysikalisches Institut Potsdam.}

\begin{abstract}
Focal ratio degradation measurements of optical fibers for the PMAS integral field instrument have 
been performed using the Photometric Testbench of the Potsdam Astrophysical Institute (AIP). An effort has been made to obtain sufficiently 
large statistical samples of individual fiber measurements.
The optimization of the measurement process has made it possible to study quantitatively the effects 
of different coupling methods (air--glass vs.\ immersion) and of imperfections like defocus, lateral 
displacement, and angular misalignment. Furthermore, the effects of mechanical stress on focal ratio
degradation have been investigated, confirming the presumed cause of modal noise which is known to limit 
the maximum signal-to-noise one can achieve with fiber-coupled echelle spectrographs.
\end{abstract}


\section{Introduction}

Optical fibers provide a versatile tool to conveniently re-arrange light
coming from different places into a common spectrograph slit. In astronomical
instrumentation, this property is being used successfully for multi-object
spectroscopy, for integral field spectroscopy, and other applications.
Similar to several new integral field spectrographs, which have been developed
recently for visual and NIR wavelengths (INTEGRAL \citep{arr98}, TEIFU \citep{hay98},
GMOS-IFU \citep{all98}, VIRMOS \citep{pri00}, IMACS-IFU \footnote{currently
under development in Durham}, CIRPASS \citep{par00}, FLAMES \citep{pas00}).
PMAS, the Potsdam Multi-Aperture Spectrophotometer \citep{rot98_2, rot98_3, rot00, lau99} was designed
as a fiber-coupled instrument, connecting a lens-array to the spectrograph slit
by means of a fiber bundle.

A basic requirement of the conceptual design consisted in a 
sufficiently stable opto-mechanical configuration which would allow for 
spectrophotometric measurements
\citep{rot97}. However, based on the observation of ``fiber noise'' in the
FOCES echelle spectrograph (M. Pfeiffer, priv. communication), there had been 
worries that the use of optical fibers might introduce instabilities, 
eventually compromising this requirement. Therefore, an optical testbench setup
was devised for measuring the far-field light distribution in the output light
cone of an optical fiber, illuminated under controlled conditions of focal
ratio, angle of incidence, centering, wavelength, and other coupling details.
By employing a cryogenic 16-bit CCD camera system, it was attempted to measure
the behaviour of the output beam pattern, both for a sufficiently
large statistical sample of fibers, as well as for the temporal behaviour of
selected fibers under geometrically stable vs.\ variable conditions. The high
dynamic range of the detector would allow to measure small differences between 
different zones (modal effects), and to determine with high precision the 
contribution of low surface brightness regions at large output angles 
(faint wings of the far field pattern), which are difficult to measure with 
simple video systems.

The far-field light distribution at the fiber output is known to exhibit, in
general, a modal structure, and to suffer from Focal Ratio Degradation (FRD), 
i.e.\ the enlargement of the output light cone angle with respect to the input beam. 
Exceeding the focal ratio of the spectrograph collimator is equivalent to light
loss and, possibly, increased stray light. Therefore, the FRD behavior of
optical fibers is an important design issue for fiber-coupled spectrographs.
FRD of a variety of fibers has been studied e.g.\ by \citep{avi88}, \citep{cra88}, 
\citep{ram88}, \citep{cla89}, \citep{car94}, \citep{avi98}. As a consequence
of the results reported in these studies, the PMAS design was chosen 
(a)~to operate with a collimator focal ratio of f/3 (the canonical value for
FRD-optimized fiber spectrographs), (b)~to avoid a bench-mounted spectrograph 
which would have required long fibers for any cassegrain-mounted IFU, and 
(c)~to employ a type of optical fiber with already demonstrated useful performance.

Because of the problem of background subtraction, the presence of FRD 
and its variability over time has rendered fiber-coupled spectrographs sometimes 
less suitable for faint object spectroscopy than conventional slit spectrographs, 
although suitable methods for calibration are capable of improving the
situation \citep{gil92}. 
We suspected that temporal variation of the fiber output far-field light 
distribution is effecting response calibrations in two ways: throughput variation 
as the result of different FRD-induced vignetting over time, and a variable 
point-spread function as the result of redistribution of light within the 
spectrograph entrance pupil due to modal effects. For IFUs, such variations must 
be expected to be independent from fiber to fiber, thus introducing calibration 
errors similar to a limited CCD flatfield accuracy. In practice, quantitative 
end-to-end measurements of such effects are difficult to perform. It was considered 
to be reasonable, however, to optimize a fiber configuration for minimal FRD 
and minimal modal redistribution if one is interested in obtaining the highest 
possible stability of the whole system.

In order to optimize the performance of the PMAS baseline design fiber and to
characterize the instrumental effects arising from fiber properties, the present 
study was conducted, including the consideration to obtain a reference for 
a future upgrade with $\approx$50 $\mu$m fibers which are required to accomodate 
a 32$\times$32 element lens array.

\section{Experimental setup}

\subsection{The Photometric Testbench}
The Photometric Testbench \citep{rot98_1} of the Potsdam Astrophysical Institute (AIP) uses several light sources and filters to create a light 
bundle with a known passband and central wavelength which passes the FRD setup. For our experiments 
a continuum white light source was used in combination with neutral density and interference filters 
(central wavelength 400$\ldots$900$\;$nm in steps of 50$\;$nm, 10$\;$nm FWHM). For most of our
measurements a central wavelength of 550$\;$nm was used. Due to the fact that the testbench is operated
under remote control from a VME controller under the VxWorks real-time operating system, and that all devices 
like filter wheels and the CCD detector can be accessed from the command line of a UNIX host computer, 
it was possible to obtain extended series of measurements automatically without human interaction.

\subsection{The FRD setup}
\begin{figure}
\plotone{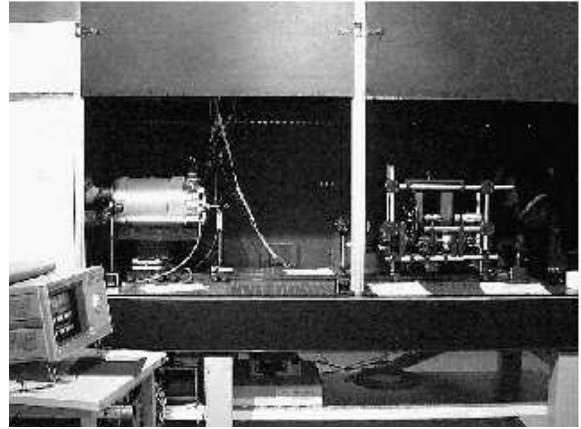}
\caption{The Photometric Testbench with the FRD setup (right) and CCD detector dewar (left).\label{testbench_foto}}
\end{figure}

\begin{figure}
\plotone{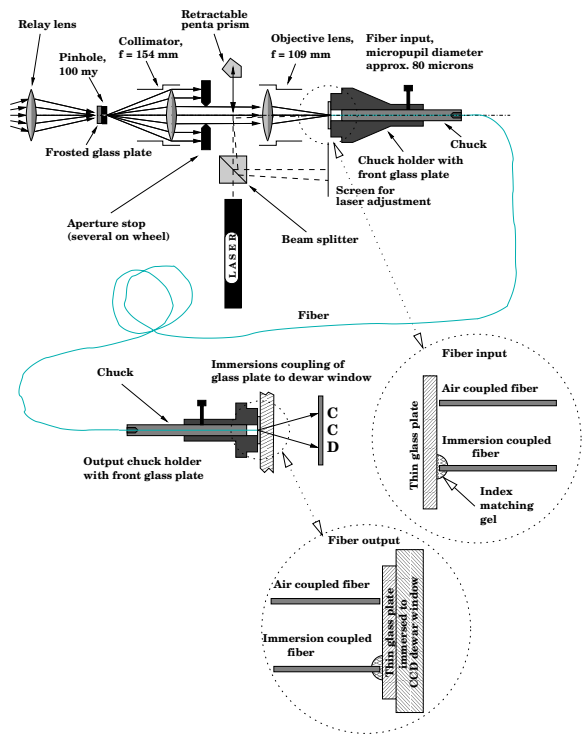}
\caption{Scheme of the optical test setup for focal ratio degradation measurements.\label{frdsetup_sketch}}
\end{figure}

The optical setup which was used for FRD measurements can be considered as part of the Photometric
Testbench and was accessible through remote control in the standard way (Fig.~\ref{testbench_foto}).
As can be seen in Fig.~\ref{frdsetup_sketch}, a pinhole behind a frosted glass plate is illuminated by 
light coming from the testbench, whose intensity and spectral passband is determined by a given 
selection of filters. The light emerging from the 100 $\mu$m - pinhole passes through a collimator 
(f = 154mm, f/1.7). The input focal ratio is determined by a motorized filter wheel with different aperture 
stops. These stops are circular, producing f-ratios ranging from f/2 to f/12. In addition, a square 
aperture is available, simulating the square lenslet input as used for PMAS.
A second lens (f = 109mm, f/1.5) focusses the light onto the fiber, which is mounted in a holder and can be 
adjusted in x, y, z, tip and tilt. The fiber itself is mounted inside of a standard Newport fiber chuck. 
For the majority of measurements the fiber was immersed to a thin glass plate using immersion gel 
(Cargille type 6307), whose index of refraction n=1.459 is very similar to quartz. 
The fiber output was prepared in a similar fashion. The fiber was immersed to an output glass plate
and the glass in turn to the window of the CCD dewar. While the FRD setup itself is in principal identical 
to those used by other authors \citep{ram88, gue88, car94}, a major advance was achieved by using a science 
grade LN$_2$-cooled CCD camera (SITe TK1024, backside illuminated, UV-visual AR-coating). This camera 
provides high sensitivity and, in particular, a high dynamic range, which is important for the precise 
measurement of intensities in the faint wings of the illuminated spot at the fiber output, as well as for 
the detection of modal noise features. 

\subsection{Measurements and data analysis}

Measurement control, data acquisition, reduction and analysis is provided automatically using the 
IDL \footnote{IDL by Research Systems Inc., Boulder, CO, USA} based code {\it PMASFIB} which was
specifically written for this application.  
After input of a list of desired focal ratios, wavelengths and light intensity levels the measurement 
cycle starts. The UNIX-workstation communicates with the CCD camera and the filter/aperture stop wheels 
through a VME controller. Several user-defined features may be invoked like pauses between exposures
(e.g.\ for testing the long-time stability of fiber setups), different angles of incidence onto the 
fiber (realized by small offsets of the aperture stop wheel), and other expandable functions. 
When the data acquisition is finished, the reduction process starts automatically. Each exposure is
background subtracted using pixels far away from the maximum spot size as determined by the nominal
fiber numerical aperture. A set of concentric circles with increasing radii is defined around the center 
of gravity. The light within these apertures is integrated and normalized to the 
total amount of light reaching the CCD. As a result, the fraction of light falling into the aperture
corresponding to the focal ratio $N_{out}$ of a virtual collimator as a function of input focal 
ratio $N_{in}$ is determined for each measurement. Also wavelength, angular input and other parameters of 
interest are recorded for each dataset. 

In what follows, the {\it f/3 coupling efficiency} $\eta_{f/3}$ is defined as fraction of light reaching a virtual f/3 collimator, normalized to the total amount of light reaching the chip. For PMAS the square lenslet input (f/3.5 along diagonal, f/5 along one side, referred to as f/3.5 $\times$ 5 in the following) was simulated by a square aperture stop of adequate size. 

\section{Improved Test Procedure}
After several first runs of an early version of the setup the advantages of immersion coupling became readily apparent. Preliminary measurements of fibers with cleaved ends were performed in unmatched and index matched versions. The observation was a considerable increase of performance of the matched fibers with respect to fibers without immersion coupling \citep{sch98}. As an explanation, it was speculated that surface irregularities due to the insufficient cleaving process had been healed out and therefore the apparent FRD was reduced significantly. Also it was observed that the variation from fiber to fiber was significantly reduced. This gave rise to investigate the behaviour on polished fibers as they were used in the final instrument.

After these preliminary tests several measures were undertaken to obtain a better degree of 
reliability for the results from FRD measurements:

\begin{itemize}
\item A laser adjustment ensured normal on-axis incidence of the input beam to the
      chuck holder front glass plate.
\item Centering of the illuminating beam on the fiber input end face was monitored by means of a photo 
      diode, connected to the fiber output.
\item The integrity of the coupling surfaces on both fiber ends was checked by observing one end with
      a microscope while illuminating the other end, and vice versa.
\item All fibers were polished and checked for visually flawless end faces.
\item A subset of fibers was measured several times using the same procedure, in order to verify
      the reliability of the process.
\item Large samples of fibers were measured.
\end{itemize}

\begin{figure}
\plotone{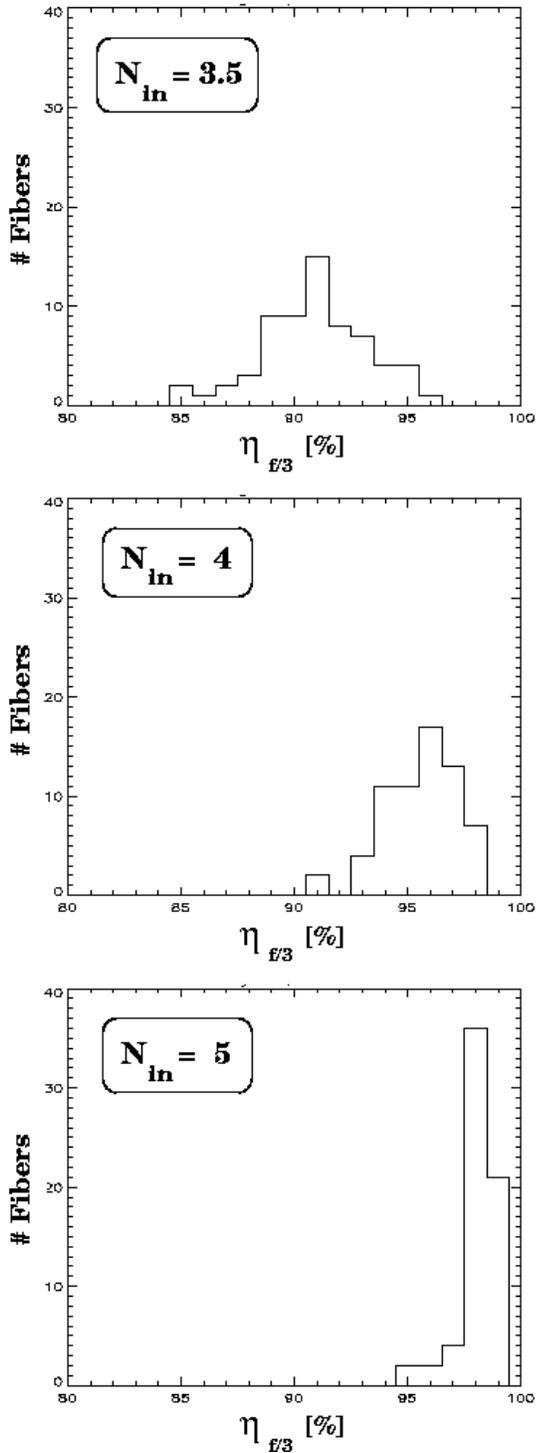}
\caption{Histograms of f/3 coupling efficiencies for input focal ratios of 3.5, 4 and 5.\label{bigsample}}
\end{figure}

Finally, a sample of 65 fibers (Polymicro FVP100/120/140) was measured using the improved FRD setup. 
The f/3 frequency of coupling efficiencies is now sharply peaked, especially when slow input focal 
ratios are used (see Fig.~\ref{bigsample}). Table \ref{bigsample_table} shows the mean and median values. 
In order to quantify the width of these distributions, $W_{100}$ and $W_{85}$ are introduced as the
the widths of these distributions, including 100 \% and 85 \% of all values around the median,
respectively.

\begin{table}
\small
\caption{$\eta_{f/3}$ - statistics for the 65 fiber sample.\label{bigsample_table}}
\begin{tabular}{ccccc}
\tableline
\tableline
$N_{in}$  & MEAN $\eta_{f/3}$ & MEDIAN $\eta_{f/3}$ & $W_{100}$ & $W_{85}$ \\
 & \% & \% & \% & \% \\
\tableline
3.5 & 90.91 & 91 & 11 & 7 \\
4 & 95.57 & 96 & 8 & 6 \\
5 & 98.11 & 98 & 5 & 3 \\
\tableline
\end{tabular}
\end{table}

To check the reliability of the measurements, three fibers from the sample above were chosen 
for ten consecutive cycles of new measurements. The choice included one with relatively low, 
one with mediocre and one with high coupling efficiency. During the measurements some conditions 
were changed (path of fibers, input and output ends, chucks). 

Table \ref{repro_table} shows that the results are not very different from 
the large sample statistics despite of the higher degree of manipulation.


\begin{table}
\small
\caption{Statistics of 30 measurements of 3 fibers.\label{repro_table}}
\begin{tabular}{ccccc}
\tableline
\tableline
$N_{in}$  & MEAN $\eta_{f/3}$ & MEDIAN $\eta_{f/3}$ & $W_{100}$ & $W_{85}$ \\
 & \% & \% & \% & \% \\
\tableline
3.5 & 91.89 & 92 & 13 & 10 \\
 4 & 95.85 & 96 & 9 & 7 \\
 5 & 98.00 & 99 & 5 & 4 \\
\tableline
\end{tabular}
\end{table}

During the tests, it became obvious that the fiber chucks, holding the fibers in a groove by means of 
a thin metal blade, have a significant effect on FRD. 

\begin{figure}
\plotone{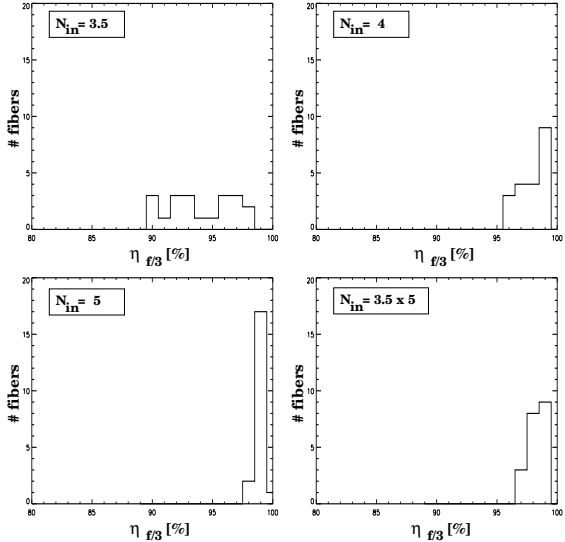}
\caption{Histograms of $\eta_{f/3}$ of 20 fibers being glued into steel tubes. $N_{in}$ = 3.5, 4, 5 and square 3.5 $\times$ 5 input. \label{ferrulensample}}
\end{figure}

Glueing the fibers into steel tubes made the results more stable, and the coupling
efficiencies increased again (see Fig.~\ref{ferrulensample} and Table \ref{ferrulensample_table})
\footnote{We applied two different types of glue, namely EPOTEK 301-2 and a very soft multi-purpose glue (UHU), which essentially produced the same results. For long-term stability, the preferred choice is the EPOTEK 301-2.}. This observation came as a surprise, because other authors have used mechanical holders precisely in order to avoid the stress normally introduced by glue \citep{avi88}. On the other hand, the effects of stress from mechanical holders was already mentioned \citep{ram88}.

\begin{table}
\small
\caption{Statistics of 20 fibers glued into steel tubes. $W_{100}$ and $W_{85}$ are the distribution widths for 100 and 85 percent of the sample. \label{ferrulensample_table}}
\begin{tabular}{ccccc}
\tableline
\tableline
$N_{in}$  & {\footnotesize MEAN} $\eta_{f/3}$ & {\footnotesize MEDIAN} $\eta_{f/3}$ & $W_{100}$ & $W_{85}$ \\
 & \% & \% & \% & \% \\
\tableline
3.5 & 94.00 & 94 & 9 & 9 \\
4 & 97.95 & 98 & 4 & 4 \\ 
5 & 98.95 & 99  & 3 & 1 \\
3.5$\times$5 & 98.30 & 98 & 3 & 3 \\
\tableline
\end{tabular}
\end{table}

\section{Final results}

\subsection{Efficiency gain by immersion coupling}

\begin{figure}
\plotone{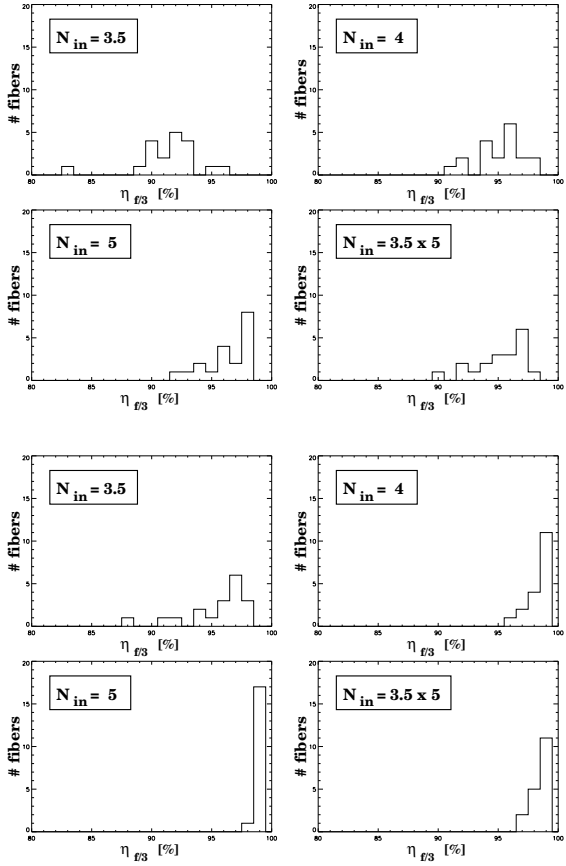}
\caption{Histograms of $\eta_{f/3}$ for $N_{in}$ = 3.5, 4, 5 and $3.5 \times 5$. Four panels on top: air coupled, bottom: immersed. \label{IMcomp_efficiencies}}
\end{figure}

\begin{figure}
\plotone{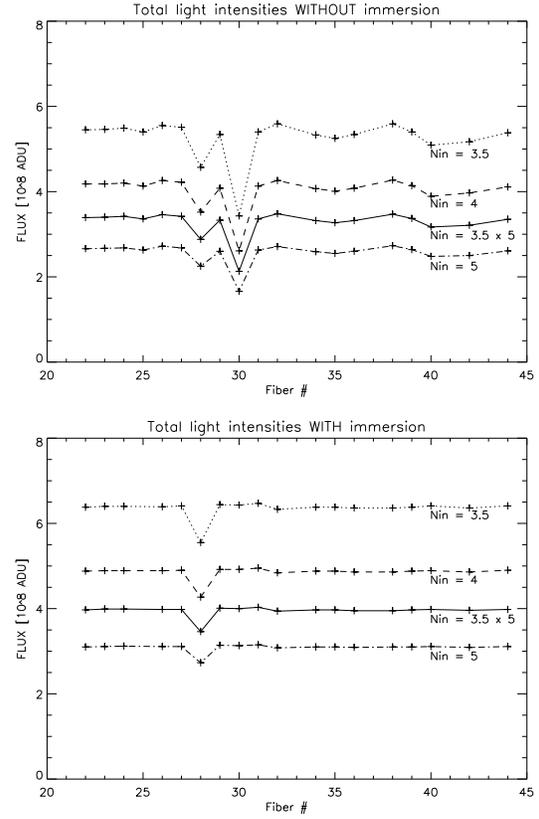}
\caption{Total throughput for air-coupled fibers (top panel) and immersed fibers (bottom).
         \label{IMcomp_intensities}}
\end{figure}

After optimizing the test setup performance, 18 of the polished fibers glued in tubes were used 
to perform a final comparison between both coupling methods. The glass plates of the chuck holders
were cleaned and the fibers attached, leaving a tiny air gap. In this way, the glass plates simulated 
the PMAS lens array on the input side, and the first collimator lens at the output. The output glass 
remained immersed to the dewar window to yield equal conditions for each test.
The data in Table~\ref{IMcomp_table} show again a larger coupling efficiency $\eta_{f/3}$ 
when immersion coupling is applied. The coupling efficency increases by nearly three percent, 
which is not as high as the observed increase of the first test. Our explanation was that this time
the fibers were not {\it cleaved}, but {\it polished}. Also the variation between different fibers 
is smaller due to polishing, but still decreases under immersion (Fig.~\ref{IMcomp_efficiencies}). 
The photometric result shows that the total amount of light $I$ increases in immersion 
(Fig.~\ref{IMcomp_intensities}). 
This was expected, because the saving of two glass-air interfaces should reduce Fresnel losses. 
Therefore, the fraction of light reaching the f/3 collimator ($I_{frac} = \eta_{f/3} I$), becomes 
significantly larger and the photometric variation between fibers decreases. The light output over all 
fibers was fairly equal, except for fiber \# 28 whose defects were not compensated by the immersion 
method. For the PMAS configuration ($N_{in} = 3.5 \times 5, N_{out}=3$) the predicted gain of light 
is 22 $\pm$ 4 \%.

\begin{table}
\footnotesize
\caption{$\eta_f/3$, total flux I as registered on CCD, and fraction of flux falling into f/3 cone $I_{frac}$ as a function of coupling method (air vs. immersion) \label{IMcomp_table}}
\begin{tabular}{cccc}
\tableline
\tableline
Method & $\eta_{f/3}$ {\tiny($W_{100};W_{85}$)} & I & $I_{frac}$ \\
 & \% & $10^8$ ADU & $10^8$ ADU \\
\tableline
AIR & 95.6 (9;6) & 3.36 $\pm$ 0.09 & 3.21 $\pm$ 0.10 \\
\tableline
OIL & 98.5 (3;2) & 3.98 $\pm$ 0.02 & 3.92 $\pm$ 0.03 \\
\tableline
\end{tabular}
\end{table}

\subsection{Influence of surface defects}

\begin{figure}
\plotone{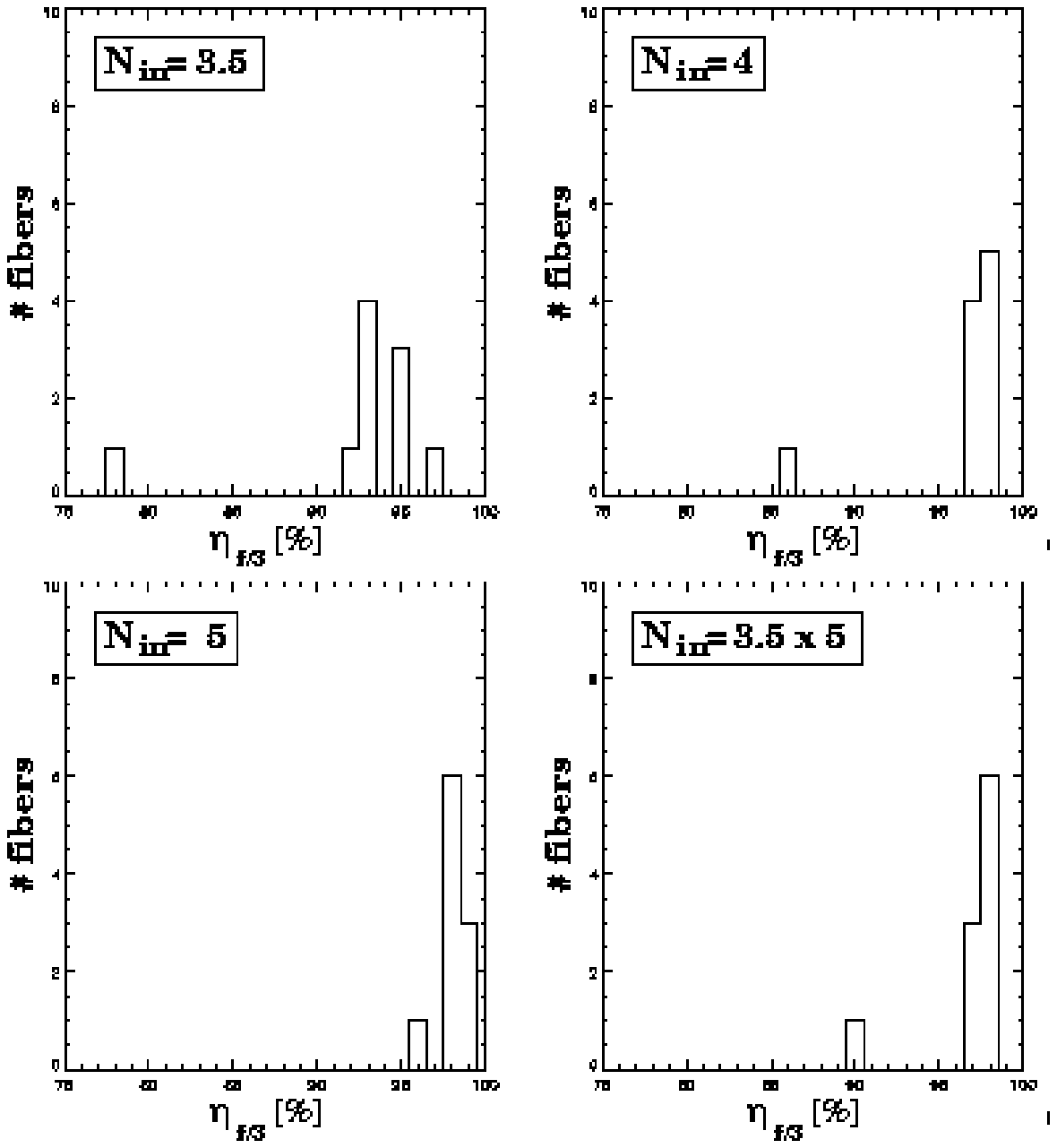}
\caption{$\eta_{f/3}$ of fibers polished down to 63 $\mu$m grain size for different input focal ratios $N_{in}$. \label{policomp_63my}}
\end{figure}

\begin{figure}
\plotone{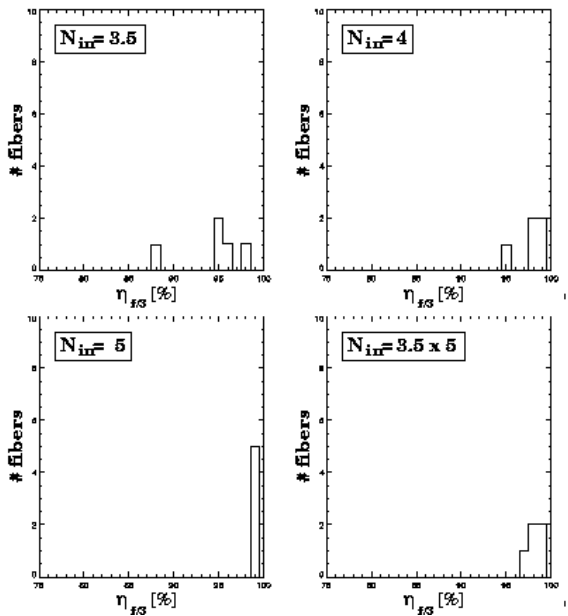}
\caption{The same as Fig.~\ref{policomp_63my} after polishing the fibers down to 9 $\mu$m grain size. \label{policomp_9my}}
\end{figure}

\begin{figure}
\plotone{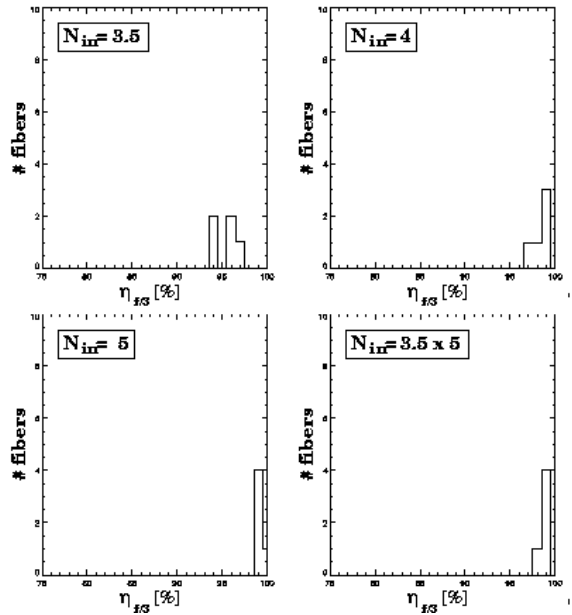}
\caption{The same as Fig.~\ref{policomp_63my} and \ref{policomp_9my}, but after polishing the fibers down to 1 $\mu$m. \label{policomp_1my}}
\end{figure}

The observed fact that surface perturbations were healed out under immersion raised the question about 
the required final fiber surface quality. To obtain an answer to this question, several fibers were
polished to different states. For the polishing process a manual polishing tool was used on abrasive 
paper with different grain sizes. Grain sizes of 63, 9, 1 and 0.3 micrometers were employed. 
As seen in Fig.~\ref{policomp_63my}, \ref{policomp_9my} and \ref{policomp_1my}, the
effect of coarse polishing is quite obvious down to 1 $\mu$m grain size. 
The statistics for 1 $\mu$m - polished fibers (Table \ref{policomp_table}) shows, albeit the poor 
statistics of only five fibers, no difference with respect to normal fibers which had been polished 
down to 0.3 $\mu$m. All these measurements were performed at a wavelength of 550 nm. 
The overall result is that high quality polishing is still necessary when using index matched fibers.

\begin{table}
\footnotesize
\caption{Dependence of $\eta_{f/3}$ from residual surface roughness \label{policomp_table}}
\begin{tabular}{ccccc}
\tableline
\tableline
Grain & $N_{in}$ & MEAN $\eta_{f/3}$ & MEDIAN $\eta_{f/3}$ & $W_{100}$ \\
$\mu$m & & \% & \% & \\
\tableline
63 & 3.5 & 92.40 & 93 & 20 \\ 
   & 4 & 96.40 & 98 & 13 \\
   & 5 & 98.10 & 98 & 4 \\
   & 3.5$\times$5 & 96.90 & 98 & 9\\
\tableline
 9 & 3.5 & 94.4 & 95 & 11\\
   & 4 & 97.8 & 98 & 5\\
   & 5 & 99.00 & 99 & 1\\
   & 3.5$\times$5 & 98.20 & 98 & 3\\
\tableline
 1 & 3.5 & 95.40 & 96 & 4\\
   & 4 & 98.40 & 99 & 3\\
   & 5 & 99.20 & 99 & 2\\
   & 3.5$\times$5 & 98.80 & 99 & 2\\
\tableline
\end{tabular}
\end{table}

\subsection{Influence of misalignment}

The input of light into fibers is critical, considering the mechanical tolerances of fiber positioning. 
The consequences of misalignment were investigated using an immersed fiber. 
The effects of defocus, translation perpendicular to the optical axis and angular misalignment have been studied.

\subsubsection{Defocus}

\begin{figure}
\plotone{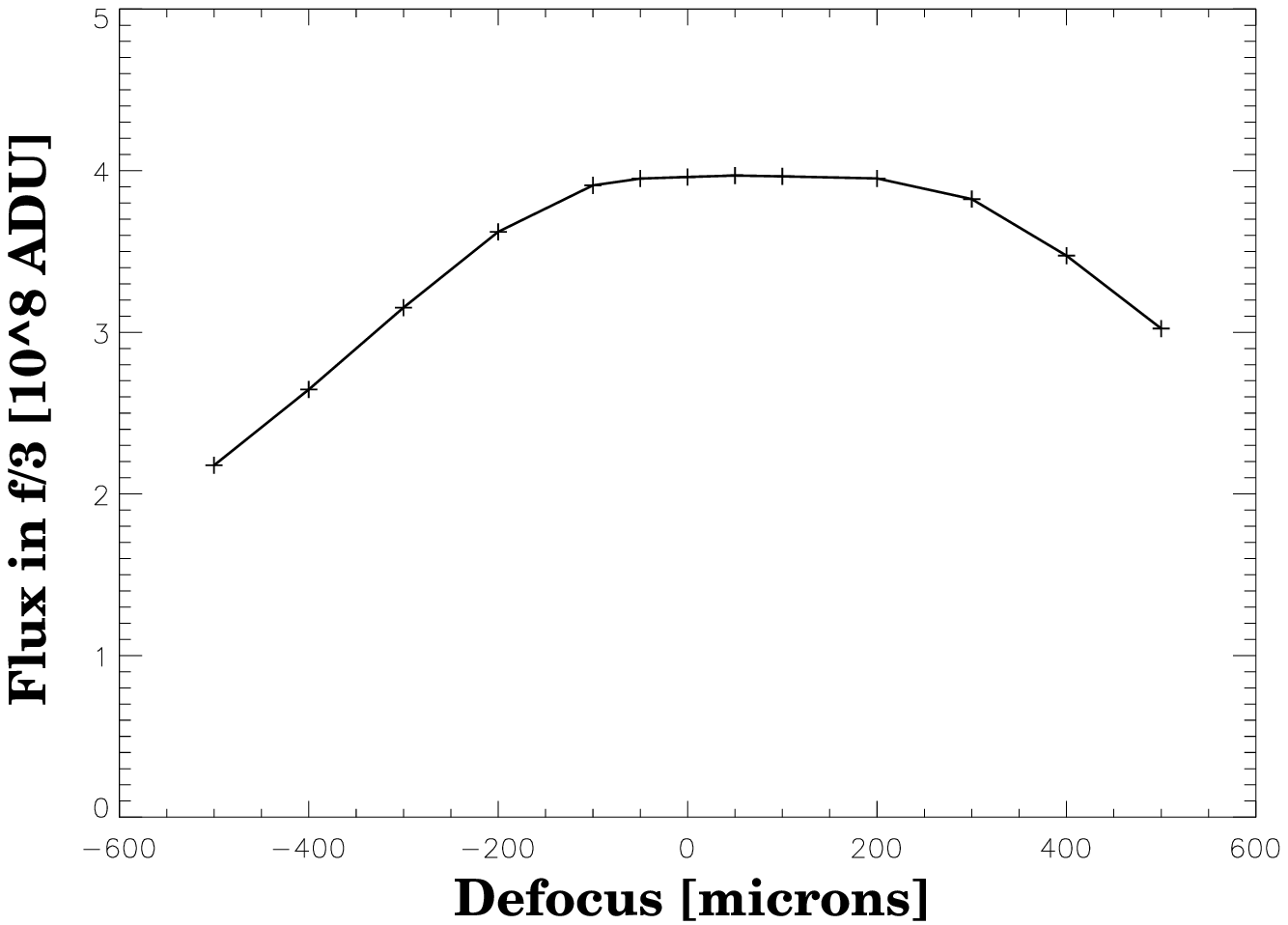}
\caption{Measured fiber output flux into f/3 as a function of defocus using square $f/3.5 \times 5$ input.\label{fiberdefocus_meas}}
\end{figure}

\begin{figure}
\plotone{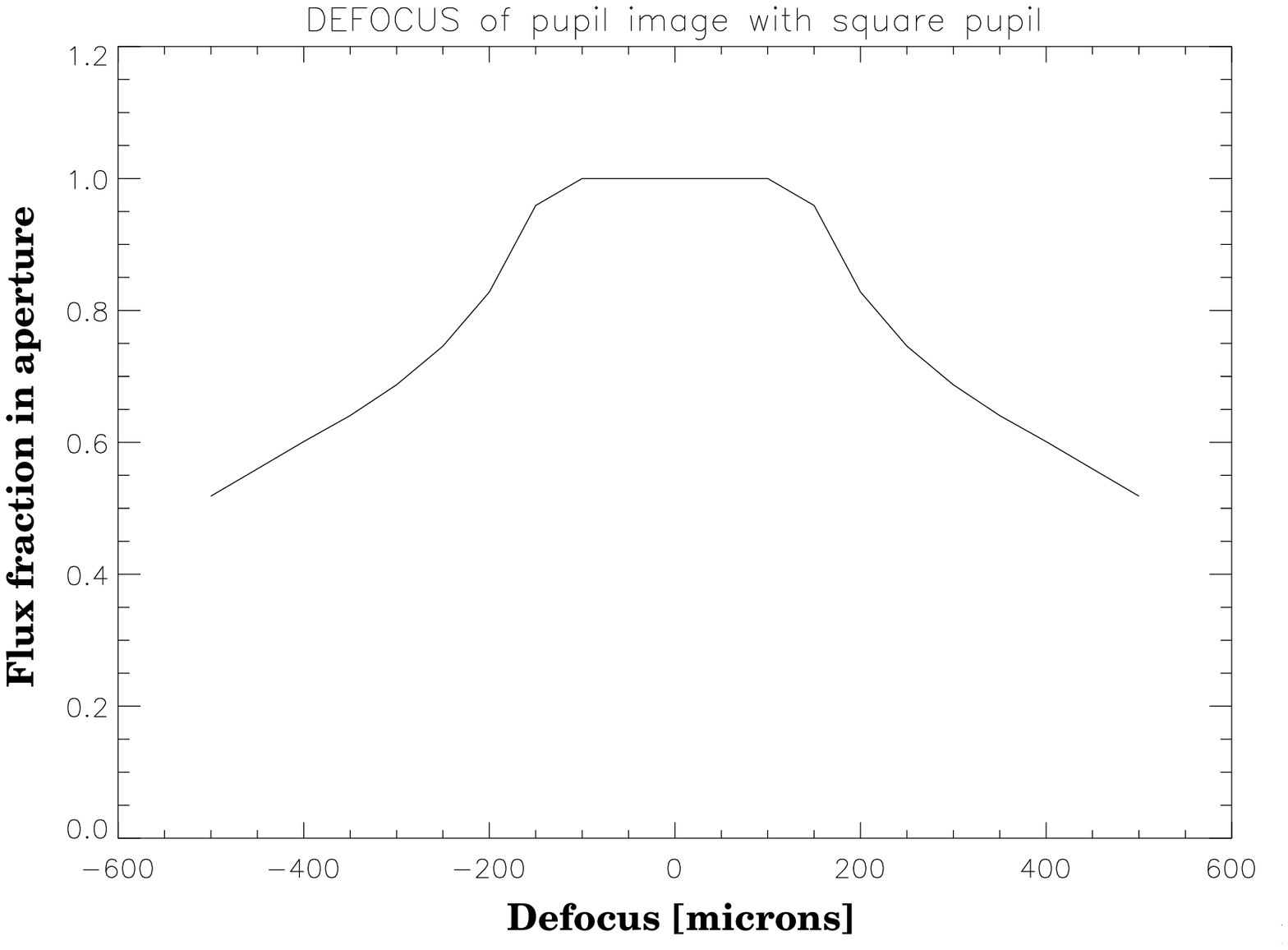}
\caption{Simulated flux into f/3 as a function of defocus, treated as pure geometrical obscuration. \label{fiberdefocus_theo}}
\end{figure}

Defocus is introduced in an immersion coupled system when the sandwich of lenslet array, substrate 
and immersion medium layer does not exactly end at the true focal plane where the fibers are attached.
Oversizing the fiber helps against errors, as seen in Fig.~\ref{fiberdefocus_meas}. 
In our example, the geometrical pupil size at the fiber input is about 71 $\mu$m. 
Some tolerance is allowed for the focal position before light is lost at the limb of the 
100 $\mu$m fiber core. While in principle FRD should not change since the angle of incidence remains 
the same, light loss sets in as soon as the tolerance of about $\pm$ 100 $\mu$m is exceeded. 
A simulation of geometric vignetting predicts fairly well the resulting light losses
(Fig.~\ref{fiberdefocus_theo}) when compared with the observed behaviour. Since diffraction and
optical aberrations were not taken into account, some details are notably different at small
displacements, but the overall picture is correct. At a defocus of 500$\mu$m, about 50\% of the
light is lost without any FRD changes.

\subsubsection{Lateral translation}

\begin{figure}
\plotone{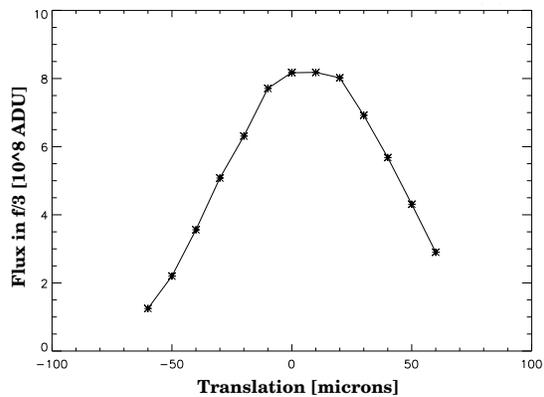}
\caption{Outgoing flux into f/3 depending on lateral shift of the input beam, $N_{in}=3.5 \times 5$.\label{fibertranslation_meas}}
\end{figure}

\begin{figure}
\plotone{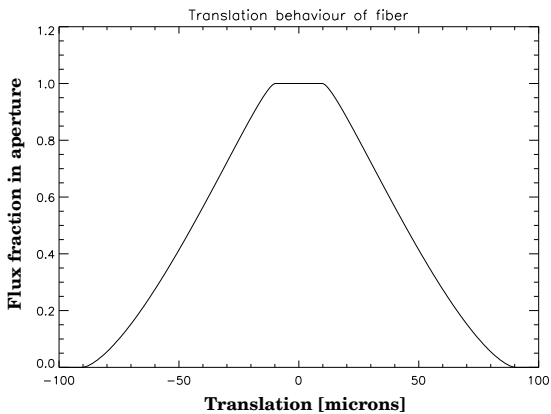}
\caption{Simulation of the situation in Fig.~\ref{fibertranslation_meas} using geometrical obscuration. \label{fibertranslation_theo}}
\end{figure}

Lateral misalignment is caused by pitch variations of the lenslet array and inaccurate fiber 
positioning. For most modern monolithic arrays, the former source of error is negligible, while
the latter remains a challenge. Errors are introduced by the mask which holds the fibers, 
the way the fibers are mounted in the mask and the degree to which the fibers are mounted
concentric to their ferrules. 
Our measurement (Fig.\ref{fibertranslation_meas}) is consistent with the allowable 
tolerance of $\pm$ 10 $\mu$m, which is obtained when a round 80 $\mu$m spot is moved inside 
a 100 $\mu$m circle as simulated in Fig.~\ref{fibertranslation_theo}. 
Note a slight misalignment of the zero point which was due to the limited accuracy of the adjustment 
with a photo diode.
The shape of the light loss behaviour is consistent with the theoretical prediction and very steep.
Even small lateral misalignments have strong effects on the overall efficiency. In this example,
more than 50\% of the light are lost as soon as the lateral misalignment is exceeding 45$\mu$m. Due 
to the fact that the fiber surface is not much larger than the image projected on, no FRD changes are 
observed.

\subsubsection{Angular misalignment}

\begin{figure}
\plotone{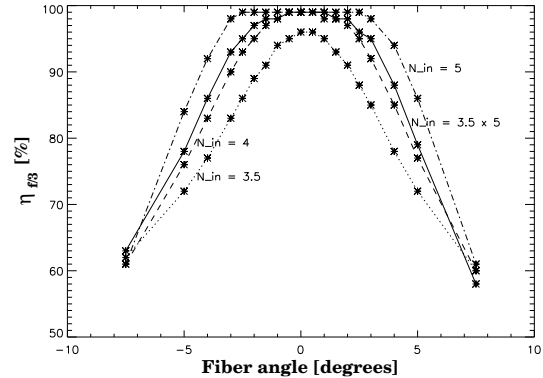}
\caption{$\eta_{f/3}$ - variation due to angular input for different $N_{in}$.\label{fiberangle}}
\end{figure}

An angular misalignment occurs when the fibers are tilted slightly in their mask or when 
the whole lenslet array is tilted during the alignment to the fore-optics. 
In our setup we simulated angular misalignment by turning the aperture stop wheel in small 
increments. Fig.~\ref{fiberangle} shows that FRD losses become significant for deviations larger than 
two degrees, which is consistent with other measurements \citep{tay93}. 
As expected, faster input focal ratios are more sensitive to angular misalignments.
In our example, a fraction of more than 30\% of the light is lost when the angular misalignment
exceeds 7 degrees. This loss is not geometric but a pure FRD loss, as the change of $\eta_{f/3}$ in 
fig. \ref{fiberangle} indicates.

\subsection{Wavelength dependence}

\begin{figure}
\plotone{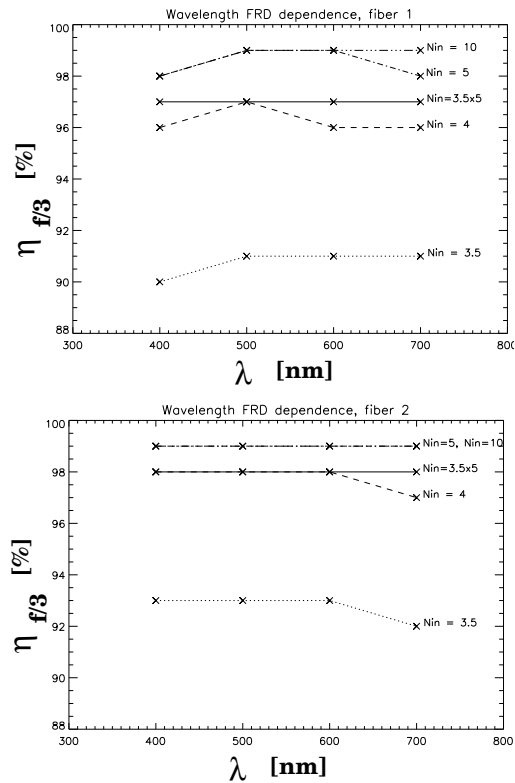}
\caption{Color dependence of $\eta_{f/3}$ for two different fibers of the same type (Polymicro FVP 100/120/140).\label{FRDcolor}}
\end{figure}

The measurements from above were performed at a wavelength of 550~nm with a bandpass of
10~nm FWHM.  A possible dependence of FRD with wavelength was investigated. 
Theoretically, no wavelength dependence is to be expected \citep{lun84}. 
Chromatic aberrations of the objective lenses resulting in focus changes with wavelength 
were measured and compensated as far as possible during the measurements. 
The results of two fibers measured (Fig.~\ref{FRDcolor}) show that the theoretical 
expectation is fulfilled in a range between 400 and 700 nm. 
For wavelengths larger then 700 nm the coupling efficiency is totally dominated by the lens
aberrations of the test setup and the results are no longer meaningful.

\subsection{Modal Noise Tests} 

\begin{figure}
\plotone{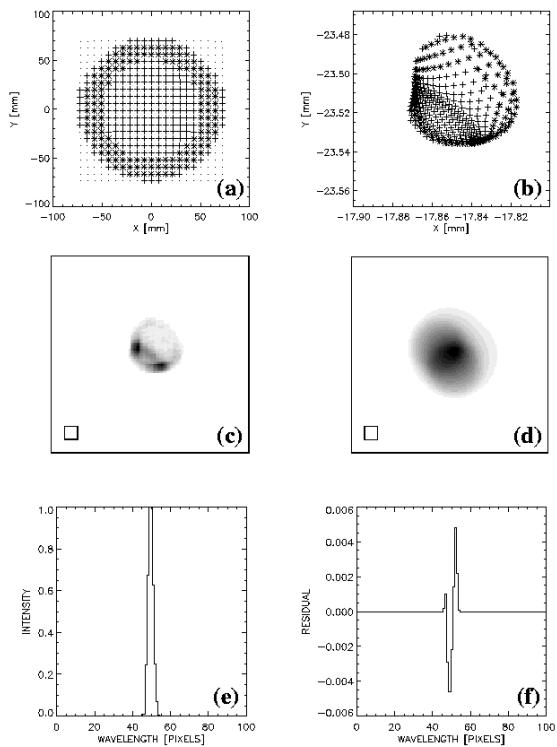}
\caption{Simulation of spectrograph PSF and the effect of modal redistribution on 
the shape of spectral line profiles. For explanation, see text. \label{SPECSIM}}
\end{figure}

In the picture of waves propagating through an optical fiber, FRD is the change
of modal distribution from lower to higher and to lossy modes. 
In bench-mounted spectrographs, connected to the telescope focal plane
with a long fiber cable, the motion of this cable during tracking may give
rise to a change of this modal distribution, caused by variable mechanical 
stress. Baudrand\&Walker (2001) have measured the effect and attributed the
observed S/N limit for high resolution spectroscopy (``modal noise'') to
the occurance of this phenomenon. The same effect had been observed to set in
with FOCES (Calar Alto 2.2m Telescope) when the telescope was tracking, but to 
vanish when the telescope stopped (M. Pfeiffer, priv. communication).

We have investigated whether modal redistribution of light in the output of
a fiber is capable of producing measurable variations of the final spectra, e.g.\ 
comparable to the typical inaccuracy of sky-subtracted deep spectra at the 
wavelength of bright night sky emission lines. We performed
a simulation, using the optical design data for the PMAS fiber spectrograph
\citep{rot02}. An example is shown in Fig.~\ref{SPECSIM} for a fiber
located at a collimator field angle of 4.1 degrees, corresponding to a position
of 82\% of the half length of the pseudo-slit, counted from the center. The 
raytracing was performed for a wavelength of 852nm in the red part of the spectrum, 
where the night sky emission lines
are known to present problems for accurate sky subtraction. Fig.~\ref{SPECSIM}a
presents the orthogonal distribution of rays in the spectrograph pupil plane, 
excluding those rays outside of the circular pupil (dots). The rays which are
entering the spectrograph focal plane (Fig.~\ref{SPECSIM}b) are plotted as
crosses and asterisks, the latter symbolizing all rays originating from
a pupil annulus whose intensity is modulated with a constant weight of (1$\pm$X\%) 
as an approximation to the observed redistribution of light in the output cone of 
a fiber. Not surprizingly,
the spot diagram (Fig.~\ref{SPECSIM}b) reveals that the annulus rays, i.e.
non-paraxial rays, are predominantly contributing to the wings of the spot,
whose image (point-spread-function, PSF) is shown in Fig.~\ref{SPECSIM}c. For 
comparison, a 15$\mu$m CCD pixel is outlined as a square in the lower left
corner or the image. The PSF shows the onset of astigmatism and coma near the 
edge  of the field. The present configuration of PMAS employs 100$\mu$m diameter
fibers, which are projected onto a demagnified diameter of 60$\mu$m (4 pixels)
on the CCD. The fiber image, represented by a circular disk of the projected
fiber diameter, was convolved with the PSF, yielding the picture shown in
Fig.~\ref{SPECSIM}d (modulation 0\%). This image was taken as a model for an unresolved 
night sky emission line, which was extracted from the simulated CCD image by coadding 
pixels along the columns. The resulting spectrum, normalized to 1, is plotted in 
Fig.~\ref{SPECSIM}e. Repeating the simulation with a pupil annulus modulation
of 10\%, and subtracting the spectrum from the unperturbed spectrum produced
the residual of Fig.~\ref{SPECSIM}f, having an amplitude of 0.5\% with respect
to the normalized emission line. The shape and order of magnitude resembles
closely the typical appearance of a less than perfectly subtracted night sky
emission line in deep spectra of faint objects. 

For simplicity, the simulation was performed along the direction of dispersion. 
We have not further investigated the effects perpendicular to the direction
of dispersion, depending largely on details like the width and spacing of 
spectra, and the method of extraction. We note qualitatively, however, that the 
expected cross-talk variance due to 10\% modal redistribution is potentially 
capable of contributing a small error of order of $\approx$1\%. 

We note also that the preponderance of non-paraxial contributions to the wings
of the PSF favours (a) refractive fiber spectrographs over optical designs with
central obstruction, and (b) a high degree of scrambling of the telescope
beam by means of a fiber, since any increased ring pattern of illumination in the 
pupil plane would magnify the effect.

\begin{figure}
\plotone{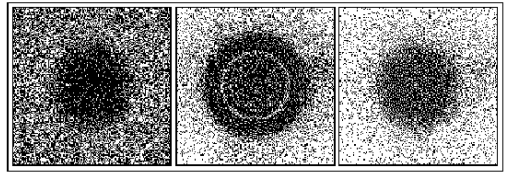}
\caption{Far field output pattern for manipulated fiber, plotted differentially to enhance contrast. 
Left: undisturbed reference, center: strong perturbation, right: weak perturbation \citep{sch98} \label{modeshow}}
\end{figure}

Our fiber testbench experiments included a stability test to verify the presence
and order of magnitude of modal redistribution under controlled conditions. 
A series of CCD images of the fiber far field output pattern was obtained and
the fiber manipulated in different ways. The test was performed under
illumination with a 550nm central wavelength, 10nm bandwidth filter.
Since the expected modulation must be observed over a large range of intensities
and is expected to be small, we chose a differential approach, 
comparing our test images with a reference frame of an undisturbed fiber. 
A low-noise, cryogenic 16-bit CCD camera is an 
ideal tool for the required accuracy of about 0.1\%. Fig.~\ref{modeshow} shows 
results from a first qualitative test, comparing an undisturbed fiber
with two cases of strong and weak perturbation (left, middle, right frame). 
Strong perturbation was imposed by pressing the fiber gently
between two fingers, while weak perturbation was the similar with
less pressure, but shaking the fiber rapidly. The static, strong stress
situation caused a modal distribution in favor of some distinct, strongly
amplified modes which created an annulus.

\begin{figure}[h]
\plotone{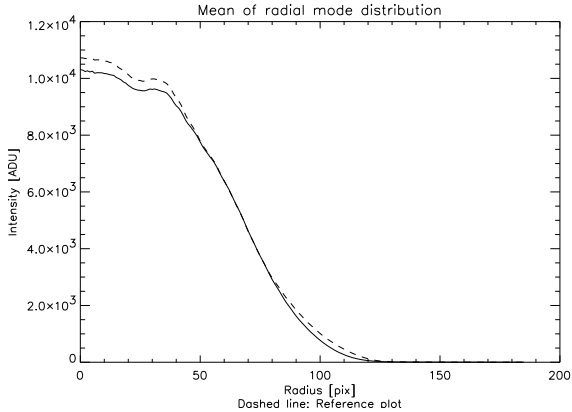}
\caption{Radial profile variations when fiber is manipulated (solid line), dashed line reference profile. Losses are observed at the profile center and outward wing.\label{mountmodes}}
\end{figure}

We also simulated the modal noise situation described by Baudrand\&Walker (2001)
with a 20m long Polymicro FHP 100/120/140 fiber and a mechanical telescope
tracking simulator, consisting in a clock driven portable amateur telescope 
mount. Over a period of 6 hours, the simulator was tracking with the fiber,
and a CCD exposure was taken every 15~minutes. Again, the series of CCD frames 
was analyzed differentially against a reference frame to observe deviations
from this reference. The most striking example is shown in Fig.~\ref{mountmodes},
revealing a drop in flux of 1.4\% near the core, and of 2.9\% in the wings,
i.e.\ a total loss of flux of 4.3\%. Neglecting the variation in intensity,
the modal redistribution is seen as a relative enhancement within a zone
similar to the annulus of our simulation from above. 

\section{Summary and Conclusions}

We have designed and built a laboratory testbench for optical fibers,
featuring a light source, filter wheels, a fiber illumination and
adjustment device, and a cryogenic, low-noise 16-bit CCD camera system.
Fiber measurements at different wavelengths and with different input
focal ratios are performed automatically under remote control from a
workstation which is also used for data reduction and analysis.

A sample of Polymicro FVP100/120/140 fibers was tested in this setup for
use in the PMAS instrument. The far field of the fiber output was recorded
with the CCD and the light collection efficiency $\eta_{f/3}$ within the 
projected nominal f/3 focal ratio of the collimator derived as a figure of 
merit for fiber FRD properties. The viability of this approach was verified
with a repeatibility test.

As a practical result, it was found that immersion coupling, glueing fibers 
into ferrules, and polishing the fiber end faces improved on average the 
f/3 light collecting efficiency, and reduced the scatter among the samples
under study.

Defocus at the fiber input was found to cause a slow decrease of $\eta_{f/3}$
beyond $\pm$~100$\mu$m. Decentering is very sensitively producing a rapid 
loss of efficiency beyond a tolerance of 15$\mu$m. Deviation of the input 
beam from normal incidence, depending on input focal ratio, caused the onset 
of losses at angles of $\pm3^\circ$ and $\pm1^\circ$\ for focal ratios of F/5 
and f/3.5, respectively. As expected, there was no wavelength dependence found 
for FRD between 400nm and 700nm. 

A qualitative test of fiber behaviour under stress showed that modal 
redistribution of light within different zones of the fiber output pattern
occured at a level of typically 1\% peak intensity for illumination at
550nm with a bandwidth of 10nm. Tracking a fiber with the telescope was 
simulated over a period of 6 hours, reproducing the effect of ``modal noise'', 
which has been discussed in the literature as the reason for a limiting
S/N for high resolution spectroscopy below the expected value due to photon
shot noise. The net effect in our test was a maximal loss of intensity of
4\%, and a redistribution of light in the output beam pattern of order several
percent. On the basis of a ray tracing model, we demonstrate that modal 
redistribution is noticeable effecting the shape of the PMAS fiber spectrograph
PSF when aberrations are present. The model is capable of reproducing typical
night sky emission line residuals.

The results of this study have entered into the design and construction of
the PMAS fiber module, but may prove useful also for other applications.
We believe that the avoidance of modal noise is mandatory for the development
of fiber coupled high resolution echelle spectrographs and spectropolarimeters
at 8-10m class telescopes, which are expected to yield very high S/N ratios up
to 10$^4$, see e.g.\citep{str03}.

\acknowledgments

This work has been supported by the German Verbundforschung under grant
05AL9BA1, and by the European Commission under contract HPRN-CT2002-00305.
The authors acknowledge the enduring support of G.\ Hasinger during all
stages of the PMAS project. The help of Evgeni Guerassimenko (Special 
Astrophysical Observatory, Zelenchuk, Russia) for the preparation and
measurement of a large amount of fibers is gratefully acknowledged.

\clearpage

\end{document}